\documentclass{svmult}
\usepackage{epsfig,graphicx} 
\usepackage[bottom]{footmisc}
\usepackage{natbib,url}      
\bibpunct{(}{)}{;}{a}{}{,}   
\def\cite#1{\citealp{#1}}    
\def\authorindex#1{}  

\begin{document}\newcount\preprintheader\preprintheader=1



\def\thisvolume{these proceedings}

\def\aj{{AJ}}			
\def\araa{{ARA\&A}}		
\def\apj{{ApJ}}			
\def\apjl{{ApJ}}		
\def\apjs{{ApJS}}		
\def\ao{{Appl.\ Optics}} 
\def\apss{{Ap\&SS}}		
\def\aap{{A\&A}}		
\def\aapr{{A\&A~Rev.}}		
\def\aaps{{A\&AS}}		
\def\an{{Astron.\ Nachrichten}}
\def\aspcs{{ASP Conf.\ Ser.}}
\def\azh{{AZh}}			
\def\baas{{BAAS}}		
\def\jrasc{{JRASC}}
\def\jgra{{JGRA}}		
\def\memras{{MmRAS}}		
\def\mnras{{MNRAS}}
\def\nat{{Nat}}		
\def\pra{{Phys.\ Rev.\ A}} 
\def\prb{{Phys.\ Rev.\ B}}		
\def\prc{{Phys.\ Rev.\ C}}		
\def\prd{{Phys.\ Rev.\ D}}		
\def\prl{{Phys.\ Rev.\ Lett}}	
\def\pasp{{PASP}}
\def\pasj{{PASJ}}		
\def\qjras{{QJRAS}}
\def\science{{Sci}}		
\def\skytel{{S\&T}}		
\def\solphys{{Solar\ Phys.}} 
\def\sovast{{Soviet\ Ast.}}  
\def\ssr{{Space\ Sci.\ Rev.}}
\def\svassp{{Astrophys.\ Space Science Proc.}}
\def\zap{{ZAp}}			
\let\astap=\aap
\let\apjlett=\apjl
\let\apjsupp=\apjs

\def\ion#1#2{{\rm #1}\,{\uppercase{#2}}}  
\def\deg{\hbox{$^\circ$}}
\def\sun{\hbox{$\odot$}}
\def\earth{\hbox{$\oplus$}}
\def\la{\mathrel{\hbox{\rlap{\hbox{\lower4pt\hbox{$\sim$}}}\hbox{$<$}}}}
\def\ga{\mathrel{\hbox{\rlap{\hbox{\lower4pt\hbox{$\sim$}}}\hbox{$>$}}}}
\def\sq{\hbox{\rlap{$\sqcap$}$\sqcup$}}
\def\arcmin{\hbox{$^\prime$}}
\def\arcsec{\hbox{$^{\prime\prime}$}}
\def\fd{\hbox{$.\!\!^{\rm d}$}}
\def\fh{\hbox{$.\!\!^{\rm h}$}}
\def\fm{\hbox{$.\!\!^{\rm m}$}}
\def\fs{\hbox{$.\!\!^{\rm s}$}}
\def\fdg{\hbox{$.\!\!^\circ$}}
\def\farcm{\hbox{$.\mkern-4mu^\prime$}}
\def\farcs{\hbox{$.\!\!^{\prime\prime}$}}
\def\fp{\hbox{$.\!\!^{\scriptscriptstyle\rm p}$}}
\def\micron{\hbox{$\mu$m}}
\def\onehalf{\hbox{$\,^1\!/_2$}}	
\def\onethird{\hbox{$\,^1\!/_3$}}
\def\twothirds{\hbox{$\,^2\!/_3$}}
\def\onequarter{\hbox{$\,^1\!/_4$}}
\def\threequarters{\hbox{$\,^3\!/_4$}}
\def\ubv{\hbox{$U\!BV$}}		
\def\ubvr{\hbox{$U\!BV\!R$}}		
\def\ubvri{\hbox{$U\!BV\!RI$}}		
\def\ubvrij{\hbox{$U\!BV\!RI\!J$}}		
\def\ubvrijh{\hbox{$U\!BV\!RI\!J\!H$}}		
\def\ubvrijhk{\hbox{$U\!BV\!RI\!J\!H\!K$}}		
\def\ub{\hbox{$U\!-\!B$}}		
\def\bv{\hbox{$B\!-\!V$}}		
\def\vr{\hbox{$V\!-\!R$}}		
\def\ur{\hbox{$U\!-\!R$}}


\def\labelitemi{{\bf --}}  

\def\rmit#1{{\it #1}}              
\def\rmit#1{{\rm #1}}              
\def\etal{\rmit{et al.}}           
\def\etc{\rmit{etc.}}           
\def\ie{\rmit{i.e.,}}              
\def\eg{\rmit{e.g.,}}              
\def\cf{cf.}                       
\def\viz{\rmit{viz.}}
\def\vs{\rmit{vs.}}

\def\rot{\hbox{\rm rot}}
\def\div{\hbox{\rm div}}
\def\lesssim{\mathrel{\hbox{\rlap{\hbox{\lower4pt\hbox{$\sim$}}}\hbox{$<$}}}}
\def\gtrsim{\mathrel{\hbox{\rlap{\hbox{\lower4pt\hbox{$\sim$}}}\hbox{$>$}}}}
\def\dif{\: {\rm d}}                       
\def\ep{\:{\rm e}^}                        
\def\dash{\hbox{$\,-\,$}}                  
\def\is{\!=\!}                             

\def\starname#1#2{${#1}$\,{\rm {#2}}}  
\def\Teff{\hbox{$T_{\rm eff}$}}   

\def\kms{\hbox{km$\;$s$^{-1}$}}
\def\Mxcm{\hbox{Mx\,cm$^{-2}$}}    

\def\Bapp{\hbox{$B_{\rm app}$}}    

\def\komega{($k, \omega$)}                 
\def\kf{($k_h,f$)}                         
\def\VminI{\hbox{$V\!\!-\!\!I$}}           
\def\IminI{\hbox{$I\!\!-\!\!I$}}           
\def\VminV{\hbox{$V\!\!-\!\!V$}}           
\def\Xt{\hbox{$X\!\!-\!t$}}                

\def\level #1 #2#3#4{$#1 \: ^{#2} \mbox{#3} ^{#4}$}   

\def\specchar#1{\uppercase{#1}}    
\def\AlI{\mbox{Al\,\specchar{i}}}  
\def\BI{\mbox{B\,\specchar{i}}} 
\def\BII{\mbox{B\,\specchar{ii}}}  
\def\BaI{\mbox{Ba\,\specchar{i}}}  
\def\BaII{\mbox{Ba\,\specchar{ii}}} 
\def\CI{\mbox{C\,\specchar{i}}} 
\def\CII{\mbox{C\,\specchar{ii}}} 
\def\CIII{\mbox{C\,\specchar{iii}}} 
\def\CIV{\mbox{C\,\specchar{iv}}} 
\def\CaI{\mbox{Ca\,\specchar{i}}} 
\def\CaII{\mbox{Ca\,\specchar{ii}}} 
\def\CaIII{\mbox{Ca\,\specchar{iii}}} 
\def\CoI{\mbox{Co\,\specchar{i}}} 
\def\CrI{\mbox{Cr\,\specchar{i}}} 
\def\CriI{\mbox{Cr\,\specchar{ii}}} 
\def\CsI{\mbox{Cs\,\specchar{i}}} 
\def\CsII{\mbox{Cs\,\specchar{ii}}} 
\def\CuI{\mbox{Cu\,\specchar{i}}} 
\def\FeI{\mbox{Fe\,\specchar{i}}} 
\def\FeII{\mbox{Fe\,\specchar{ii}}} 
\def\FeIX{\mbox{Fe\,\specchar{ix}}}
\def\FeX{\mbox{Fe\,\specchar{x}}}
\def\FeXVI{\mbox{Fe\,\specchar{xvi}}}
\def\FrI{\mbox{Fr\,\specchar{i}}}
\def\HI{\mbox{H\,\specchar{i}}} 
\def\HII{\mbox{H\,\specchar{ii}}} 
\def\Hmin{\hbox{\rmH$^{^{_{\scriptstyle -}}}$}}      
\def\Hemin{\hbox{{\rm He}$^{^{_{\scriptstyle -}}}$}} 
\def\HeI{\mbox{He\,\specchar{i}}} 
\def\HeII{\mbox{He\,\specchar{ii}}} 
\def\HeIII{\mbox{He\,\specchar{iii}}} 
\def\KI{\mbox{K\,\specchar{i}}} 
\def\KII{\mbox{K\,\specchar{ii}}} 
\def\KIII{\mbox{K\,\specchar{iii}}} 
\def\LiI{\mbox{Li\,\specchar{i}}} 
\def\LiII{\mbox{Li\,\specchar{ii}}} 
\def\LiIII{\mbox{Li\,\specchar{iii}}} 
\def\MgI{\mbox{Mg\,\specchar{i}}} 
\def\MgII{\mbox{Mg\,\specchar{ii}}} 
\def\MgIII{\mbox{Mg\,\specchar{iii}}} 
\def\MnI{\mbox{Mn\,\specchar{i}}} 
\def\NI{\mbox{N\,\specchar{i}}}
\def\NaI{\mbox{Na\,\specchar{i}}}
\def\NaII{\mbox{Na\,\specchar{ii}}}
\def\NaIII{\mbox{Na\,\specchar{iii}}} 
\def\NiI{\mbox{Ni\,\specchar{i}}} 
\def\NiII{\mbox{Ni\,\specchar{ii}}}
\def\NiIII{\mbox{Ni\,\specchar{iii}}} 
\def\OI{\mbox{O\,\specchar{i}}} 
\def\OVI{\mbox{O\,\specchar{vi}}}
\def\RbI{\mbox{Rb\,\specchar{i}}} 
\def\SII{\mbox{S\,\specchar{ii}}} 
\def\SiI{\mbox{Si\,\specchar{i}}} 
\def\SiII{\mbox{Si\,\specchar{ii}}} 
\def\SrI{\mbox{Sr\,\specchar{i}}}
\def\SrII{\mbox{Sr\,\specchar{ii}}}
\def\TiI{\mbox{Ti\,\specchar{i}}} 
\def\TiII{\mbox{Ti\,\specchar{ii}}} 
\def\TiIII{\mbox{Ti\,\specchar{iii}}} 
\def\TiIV{\mbox{Ti\,\specchar{iv}}} 
\def\VI{\mbox{V\,\specchar{i}}} 
\def\HtwoO{\mbox{H$_2$O}}        
\def\Otwo{\mbox{O$_2$}}          

\def\Halpha{\mbox{H\hspace{0.1ex}$\alpha$}} 
\def\Ha{\mbox{H\hspace{0.2ex}$\alpha$}}
\def\Hbeta{\mbox{H\hspace{0.2ex}$\beta$}}
\def\Hgamma{\mbox{H\hspace{0.2ex}$\gamma$}}
\def\Hdelta{\mbox{H\hspace{0.2ex}$\delta$}}
\def\Hepsilon{\mbox{H\hspace{0.2ex}$\epsilon$}}
\def\Hzeta{\mbox{H\hspace{0.2ex}$\zeta$}}
\def\Lyalpha{\mbox{Ly$\hspace{0.2ex}\alpha$}}
\def\Lybeta{\mbox{Ly$\hspace{0.2ex}\beta$}}
\def\Lygamma{\mbox{Ly$\hspace{0.2ex}\gamma$}}
\def\Lycont{\mbox{Ly\hspace{0.2ex}{\small cont}}}
\def\Baalpha{\mbox{Ba$\hspace{0.2ex}\alpha$}}
\def\Babeta{\mbox{Ba$\hspace{0.2ex}\beta$}}
\def\Bacont{\mbox{Ba\hspace{0.2ex}{\small cont}}}
\def\Paalpha{\mbox{Pa$\hspace{0.2ex}\alpha$}}
\def\Bralpha{\mbox{Br$\hspace{0.2ex}\alpha$}}

\def\NaD{\mbox{Na\,\specchar{i}\,D}}    
\def\NaDone{\mbox{Na\,\specchar{i}\,\,D$_1$}}
\def\NaDtwo{\mbox{Na\,\specchar{i}\,\,D$_2$}}
\def\NaID{\mbox{Na\,\specchar{i}\,\,D}}
\def\NaIDone{\mbox{Na\,\specchar{i}\,\,D$_1$}}
\def\NaIDtwo{\mbox{Na\,\specchar{i}\,\,D$_2$}}
\def\Done{\mbox{D$_1$}}
\def\Dtwo{\mbox{D$_2$}}

\def\Mgbone{\mbox{Mg\,\specchar{i}\,b$_1$}}
\def\Mgbtwo{\mbox{Mg\,\specchar{i}\,b$_2$}}
\def\Mgbthree{\mbox{Mg\,\specchar{i}\,b$_3$}}
\def\MgIb{\mbox{Mg\,\specchar{i}\,b}}
\def\MgIbone{\mbox{Mg\,\specchar{i}\,b$_1$}}
\def\MgIbtwo{\mbox{Mg\,\specchar{i}\,b$_2$}}
\def\MgIbthree{\mbox{Mg\,\specchar{i}\,b$_3$}}

\def\CaIIK{\mbox{Ca\,\specchar{ii}\,K}}       
\def\CaIIH{\mbox{Ca\,\specchar{ii}\,H}}
\def\CaIIHK{\mbox{Ca\,\specchar{ii}\,H\,\&\,K}}
\def\HK{\mbox{H\,\&\,K}}
\def\Kthree{\mbox{K$_3$}}      
\def\Hthree{\mbox{H$_3$}}
\def\Ktwo{\mbox{K$_2$}}
\def\Htwo{\mbox{H$_2$}}
\def\Kone{\mbox{K$_1$}}     
\def\Hone{\mbox{H$_1$}}     
\def\KtwoV{\mbox{K$_{2V}$}}
\def\KtwoR{\mbox{K$_{2R}$}}
\def\KoneV{\mbox{K$_{1V}$}}
\def\KoneR{\mbox{K$_{1R}$}}
\def\HtwoV{\mbox{H$_{2V}$}}
\def\HtwoR{\mbox{H$_{2R}$}}
\def\HoneV{\mbox{H$_{1V}$}}
\def\HoneR{\mbox{H$_{1R}$}}

\def\hk{\mbox{h\,\&\,k}}
\def\kthree{\mbox{k$_3$}}    
\def\hthree{\mbox{h$_3$}}
\def\ktwo{\mbox{k$_2$}}
\def\htwo{\mbox{h$_2$}}
\def\kone{\mbox{k$_1$}}     
\def\hone{\mbox{h$_1$}}     
\def\ktwoV{\mbox{k$_{2V}$}}
\def\ktwoR{\mbox{k$_{2R}$}}
\def\koneV{\mbox{k$_{1V}$}}
\def\koneR{\mbox{k$_{1R}$}}
\def\htwoV{\mbox{h$_{2V}$}}
\def\htwoR{\mbox{h$_{2R}$}}
\def\honeV{\mbox{h$_{1V}$}}
\def\honeR{\mbox{h$_{1R}$}}
  

\title*{Multi-Wavelength View of Flare Events on November 20, 2003}

\titlerunning{Flare Events on November 20, 2003} 

\author{P. Kumar\inst{1}
        \and
         P. K. Manoharan\inst{2}
        \and
         W. Uddin\inst{1}}


\authorindex{Kumar, P.}
\authorindex{Manoharan, P. K.}
\authorindex{Uddin, W.}


\authorrunning{Kumar et al.}  

\institute{Aryabhatta Research Institute of Observational Sciences, Nainital, India
           \and 
 Radio Astronomy Centre (NCRA), Tata Institute of Fundamental Research, Udhagamandalam (Ooty), India}


\maketitle

\setcounter{footnote}{0}  

\begin{abstract} 
We analyze two flare events which occurred in active region NOAA 501
on November 20, 2003. The H$\alpha$ and magnetogram measurements show
interaction between two filaments which produced a slowly rising flare
event, corresponding to two stages of magnetic reconnection. The
relative clockwise rotation between the two sunspot systems caused
filament destabilization. The cusp-shaped magnetic field in the main
phase of the second flare and its evolution in correlation with ribbon
separation provide evidence for the cause of the CME eruption. The
propagation and orientation of the CME with respect to the ecliptic
plane is illustrated by IPS images.
  
\end{abstract}


\section{Flare events on November 20, 2003}      \label{kumar-sec: Flare Events on 20 November 2003}

The period of October--November 2003 is well known for its extreme
solar activity, corresponding to ARs 484, 486 and 488
(\cite{kumar-2006JAA...27..255U}).  The continuous emergence of
magnetic flux in and around AR 501 (return of AR484, in the next
rotation) also caused several intense flares. Here, we report two
flare events (1N/M1.4 and 2B/M9.6) that occurred near N00\,W05 on
November 20, 2003.  GOES soft X-ray measurements at 0.5--4\,\AA\ and
1--8\,\AA\ bands (Fig. 1) show a broad profile of gradual rise and decline
for the event at 02:12 UT.  However, the second event (peak at 07:47
UT) showed a quick rise in intensity and a gradual decrease.  
Before the start of the latter event, an impulsive flare (C3.8) was
observed at 07:25 UT.

\begin{figure}
\centering
\includegraphics[width=7.0cm]{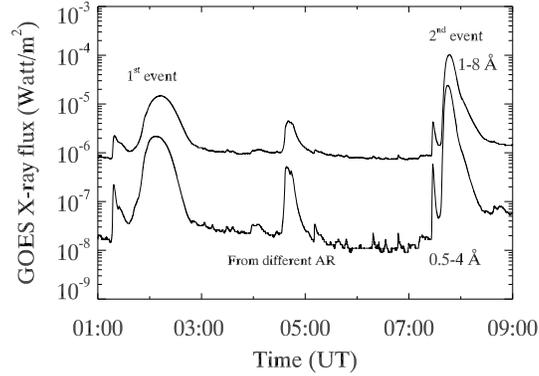}
\caption []{\label{kumar-fig:1}
GOES X-ray measurements in the 0.5--4\,\AA\ and 1--8\,\AA\ wavelength bands.}
\end{figure}
\section{Re-organization of magnetic structures}    \label{kumar-sec: Re-organisation of Magnetic Structures}

H$\alpha$ images were recorded at ARIES, Nainital, with a cadence of
about 15--20\,s (details in \cite{kumar-2007SoPh..242..143}).  The
high sampling rate allows us to follow the evolution of the complex
twisted field structures in the chromosphere
in relataion to the magnetic configuration in the photoshere 
seen on MDI images (Fig.~ 2). 
Before the onset of the first flare, a system of two filaments goes
through a gradual evolution (Fig.~2 a-c, indicated by arrows) while
they also approach each other at a speed of about 10~\kms\
(Fig.~3). The interaction between them triggers the energy release,
during 01:50--02:10~UT and leads to field merging seen in the
H$\alpha$ images. The change in filament orientation after the energy
release indicates a relaxed state of the magnetic configuration.

\begin{figure}

\centering
\includegraphics[width=2.00cm]{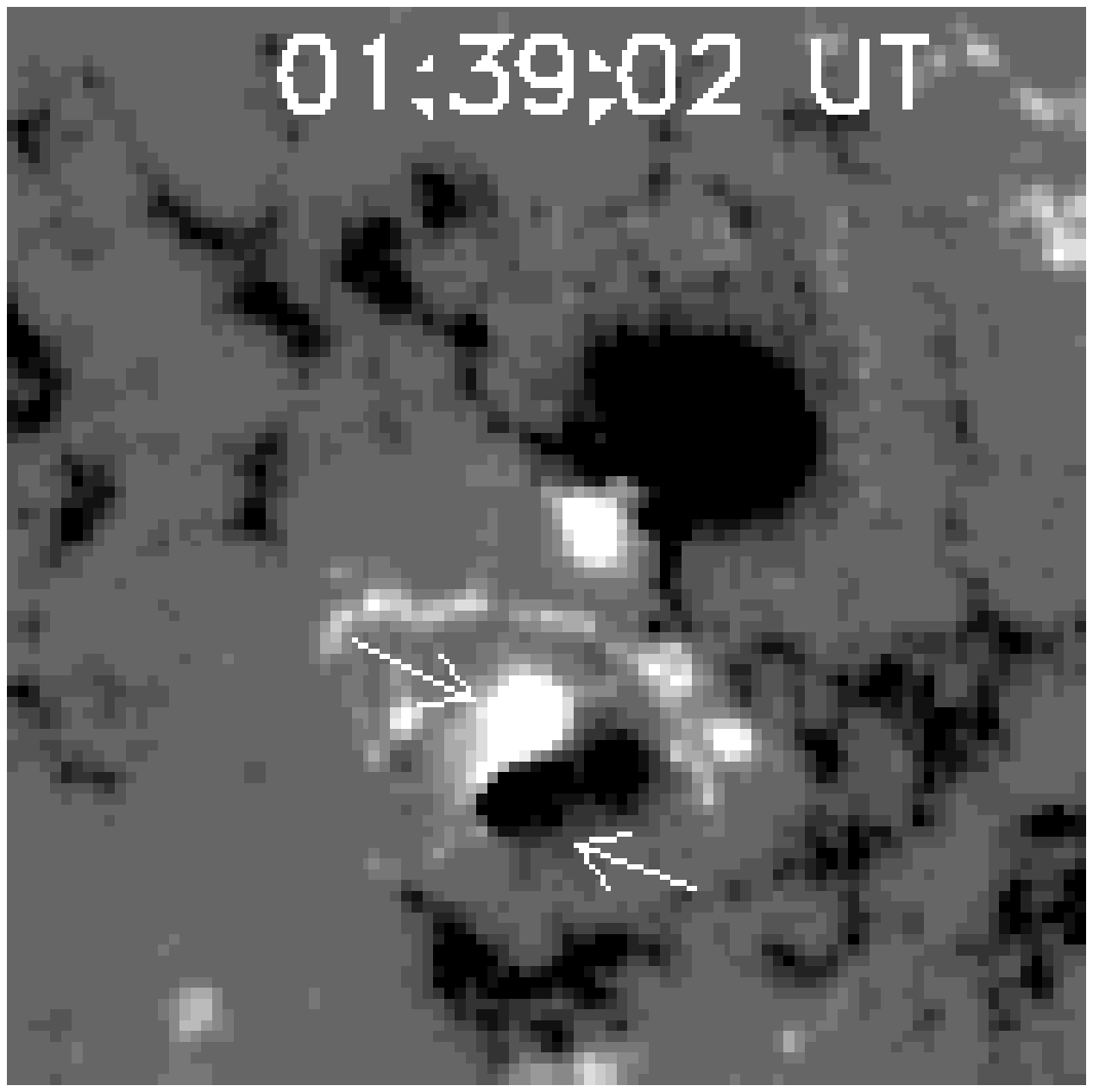}
\includegraphics[width=2.00cm]{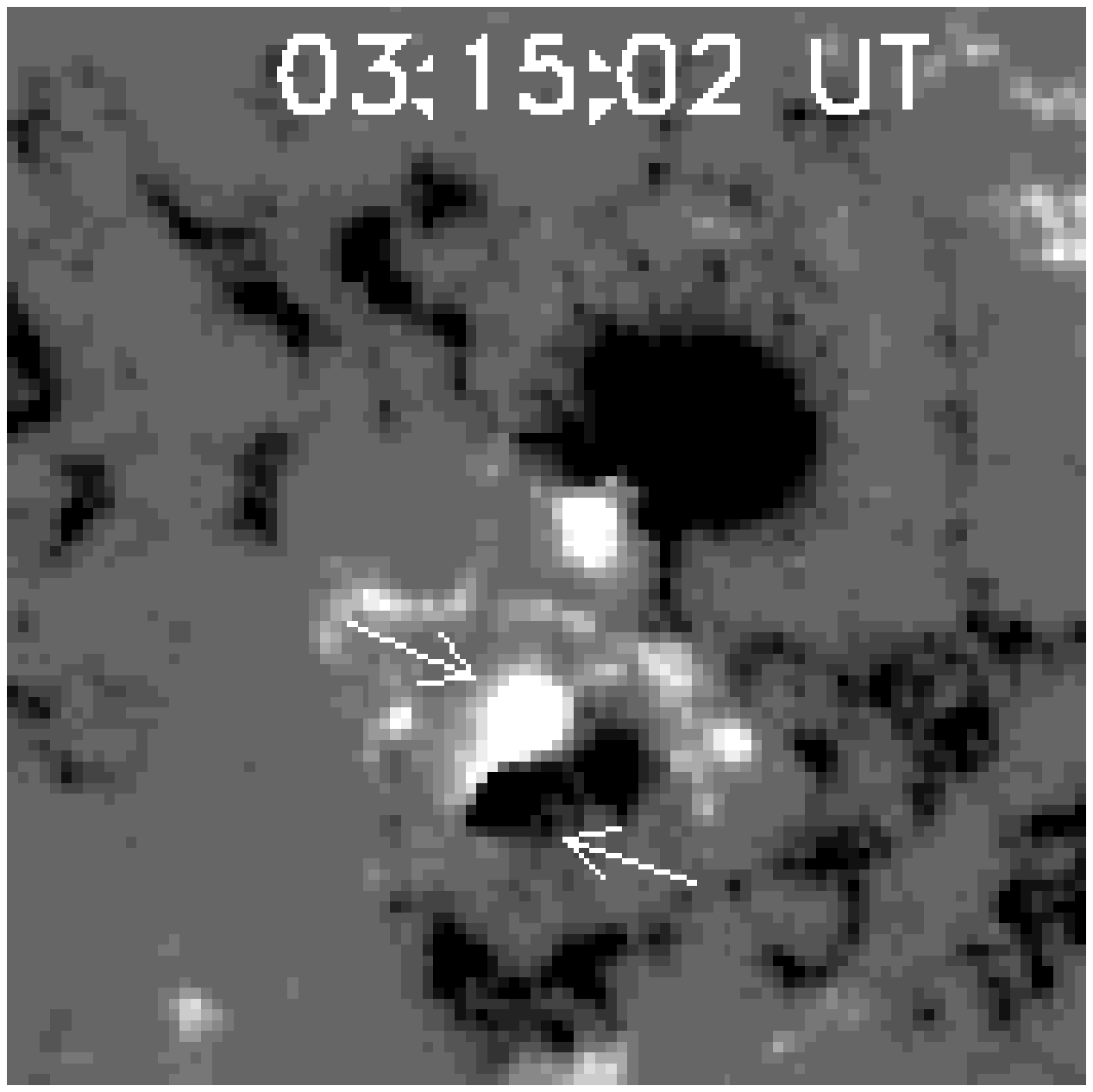}
\includegraphics[width=2.00cm]{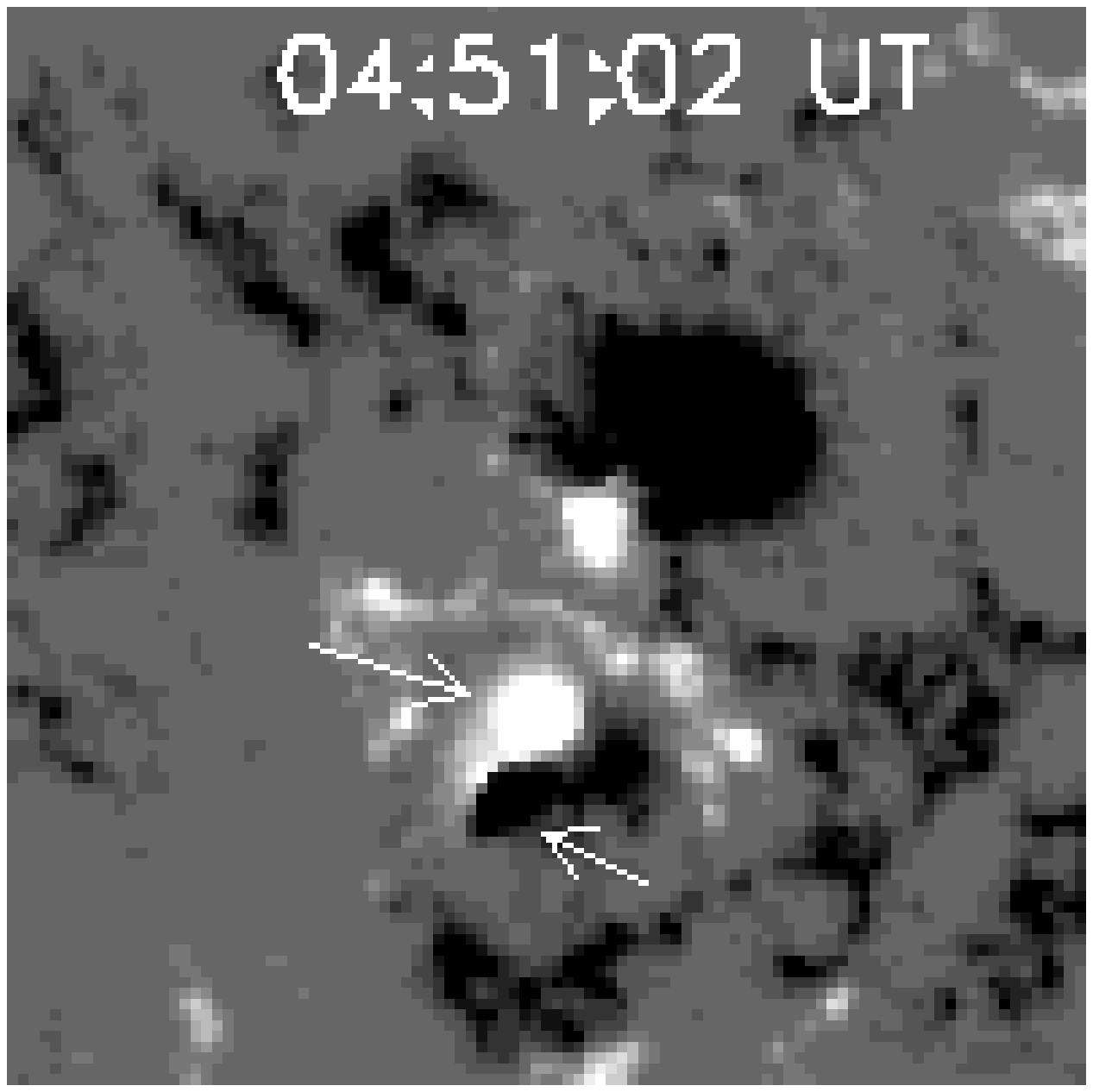}
\includegraphics[width=2.00cm]{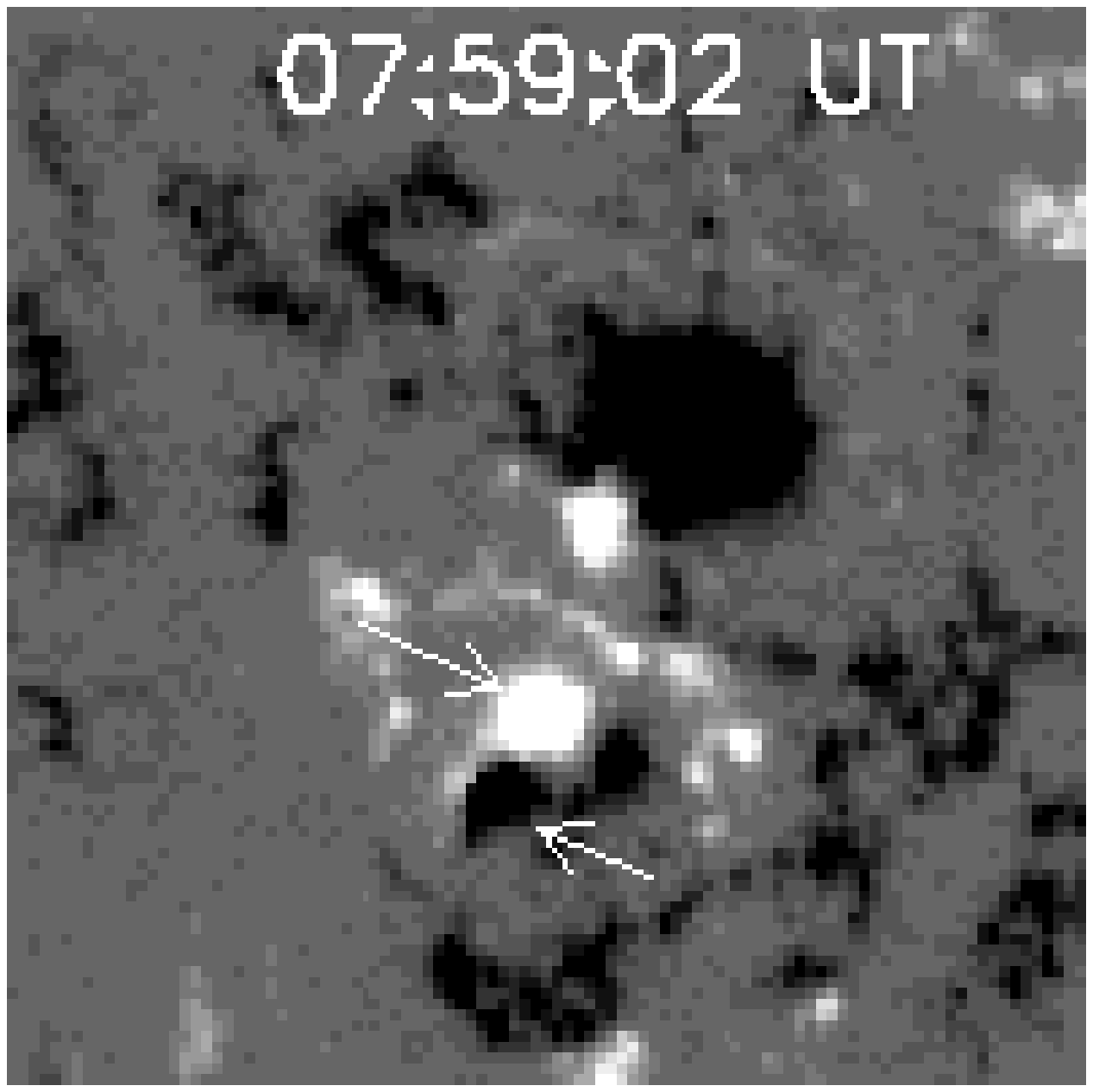}

\includegraphics[width=2.00cm]{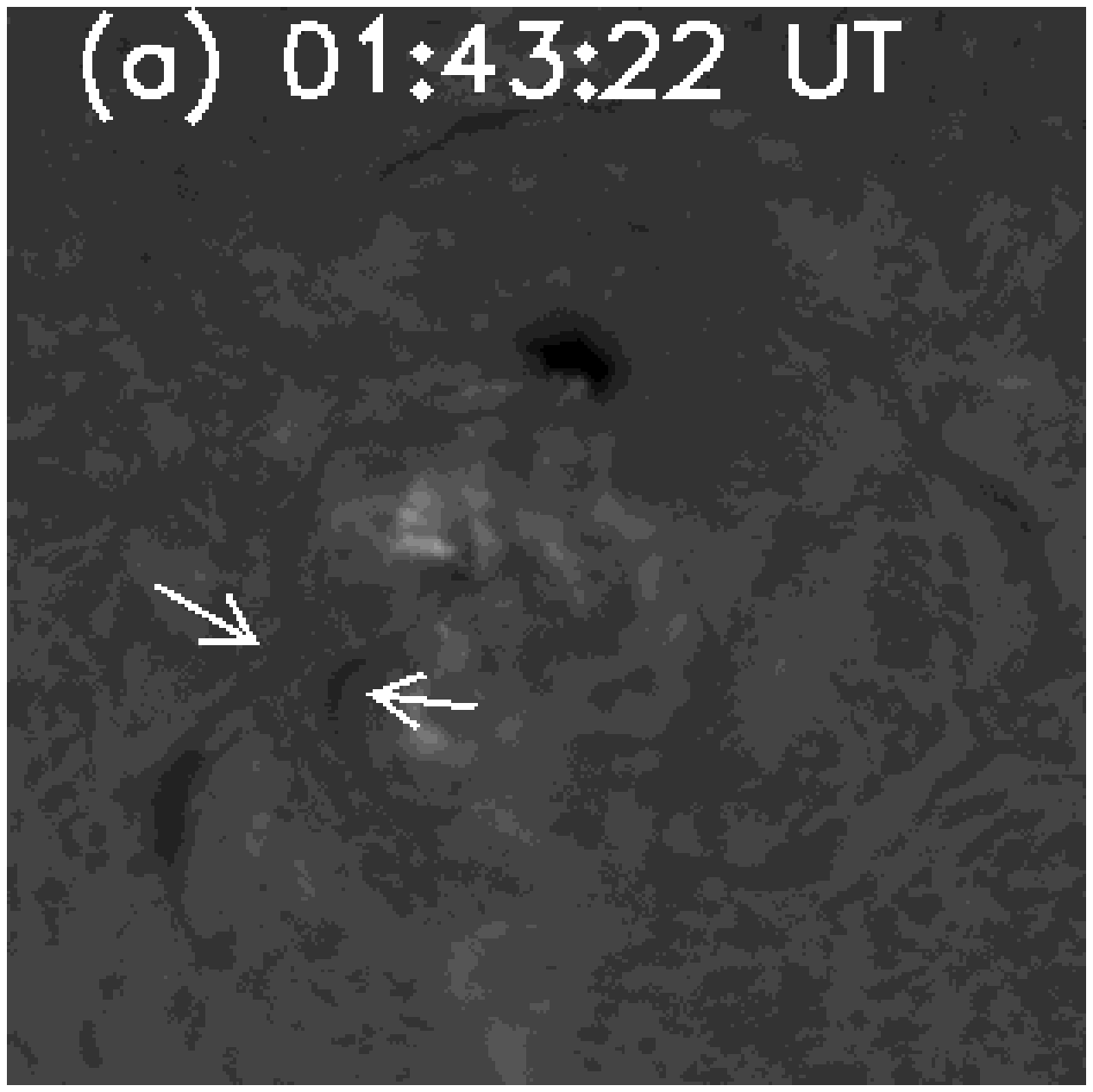}
\includegraphics[width=2.00cm]{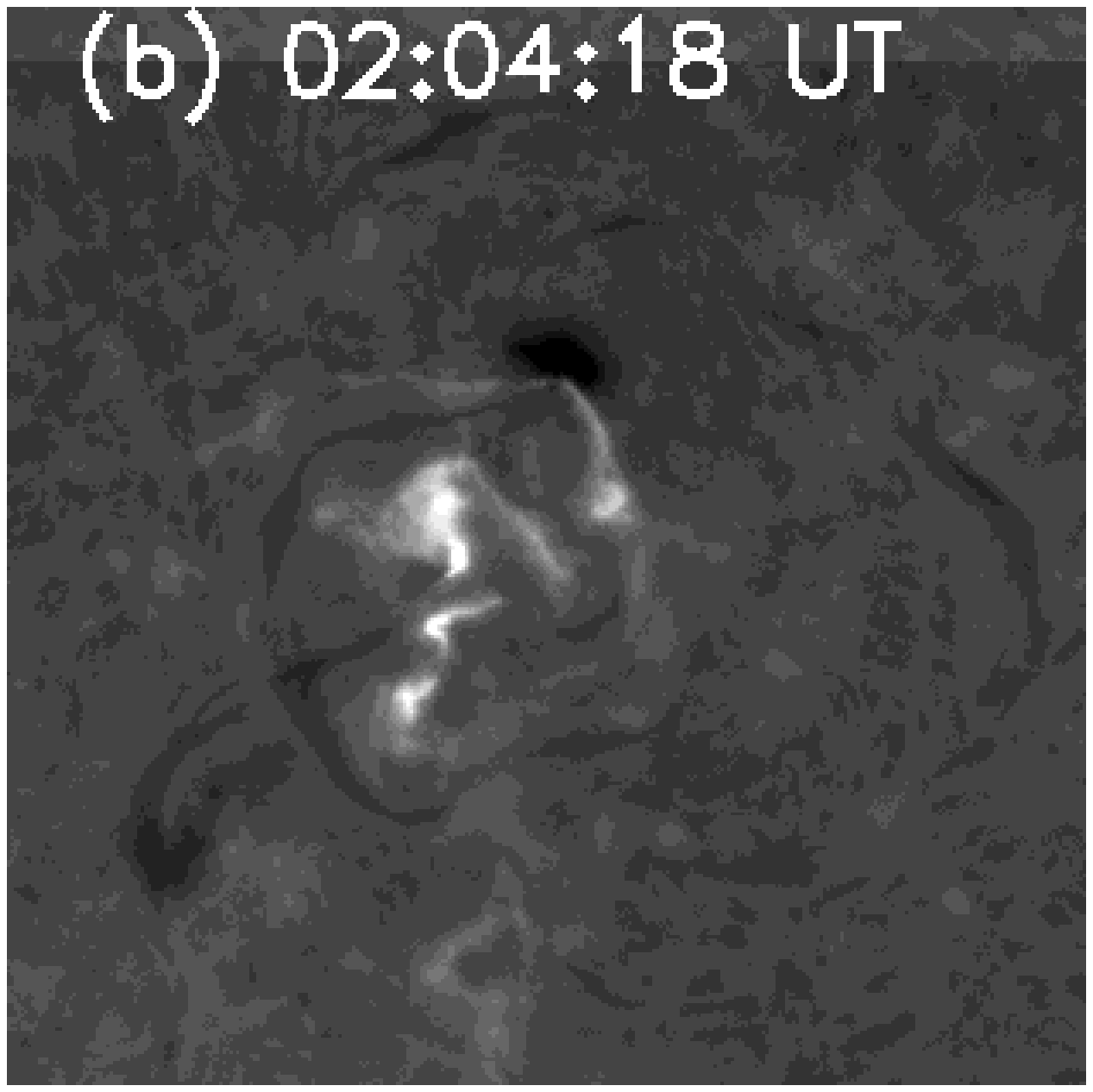}
\includegraphics[width=2.00cm]{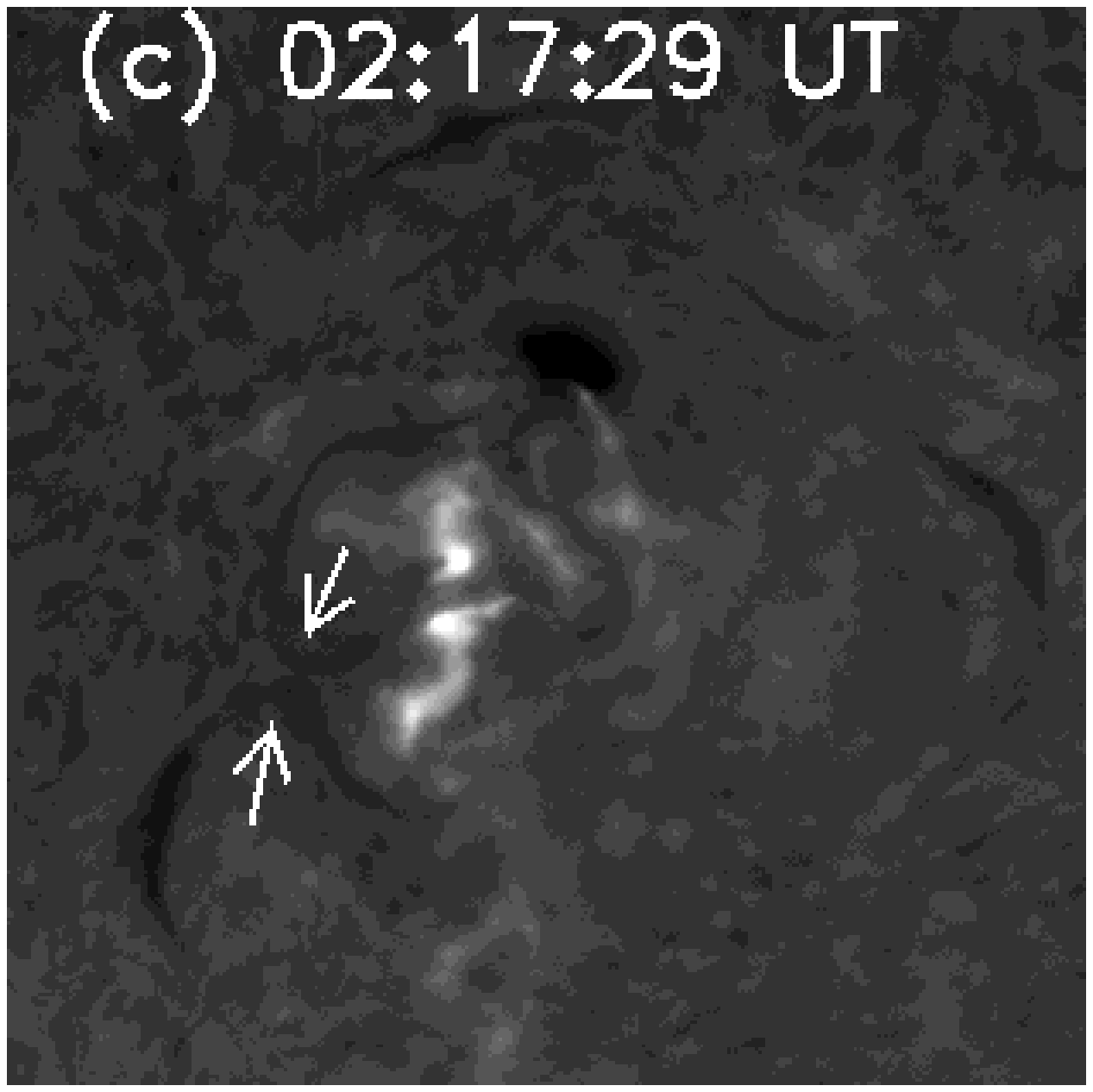}
\includegraphics[width=2.00cm]{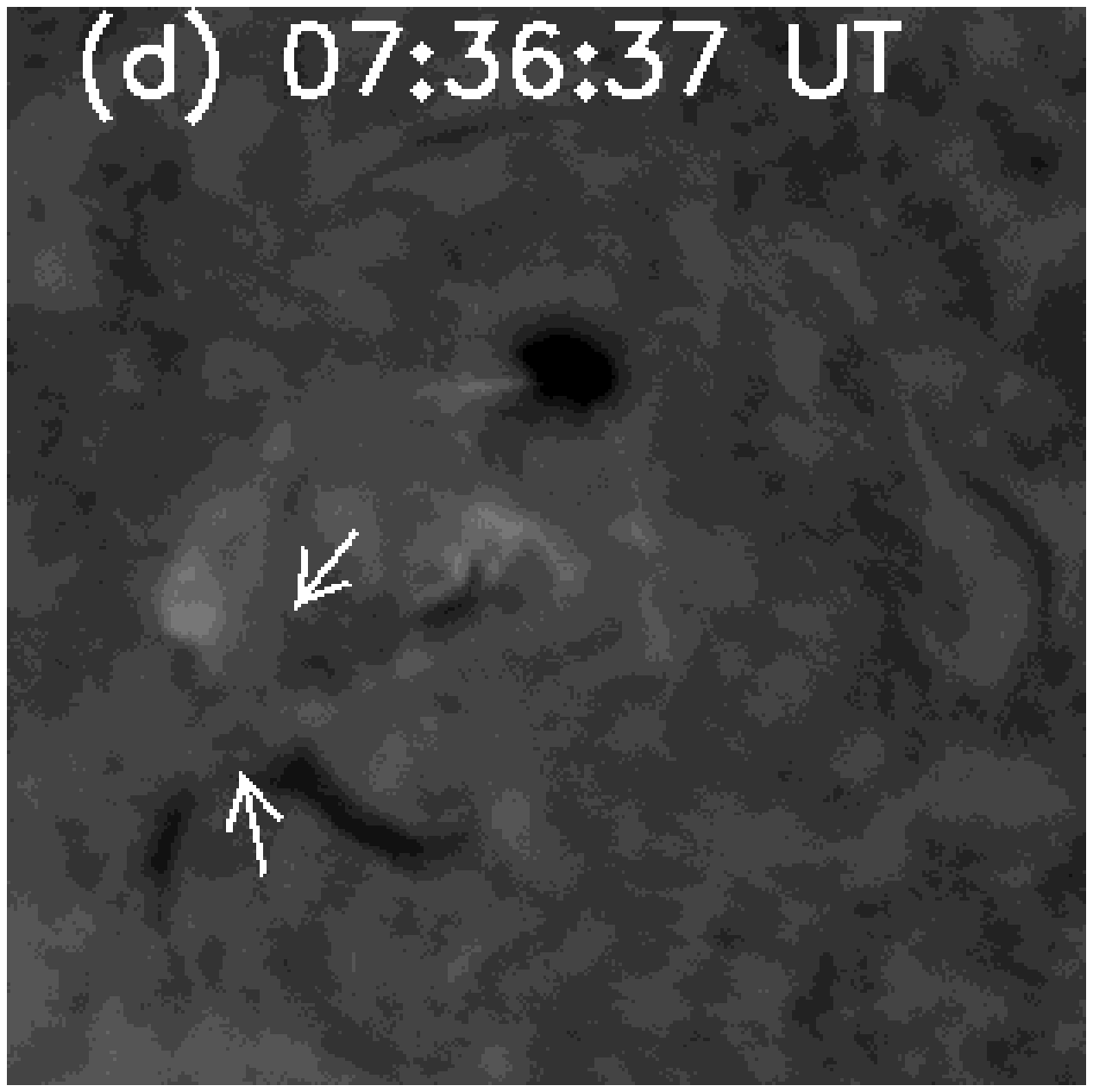}

\includegraphics[width=2.00cm]{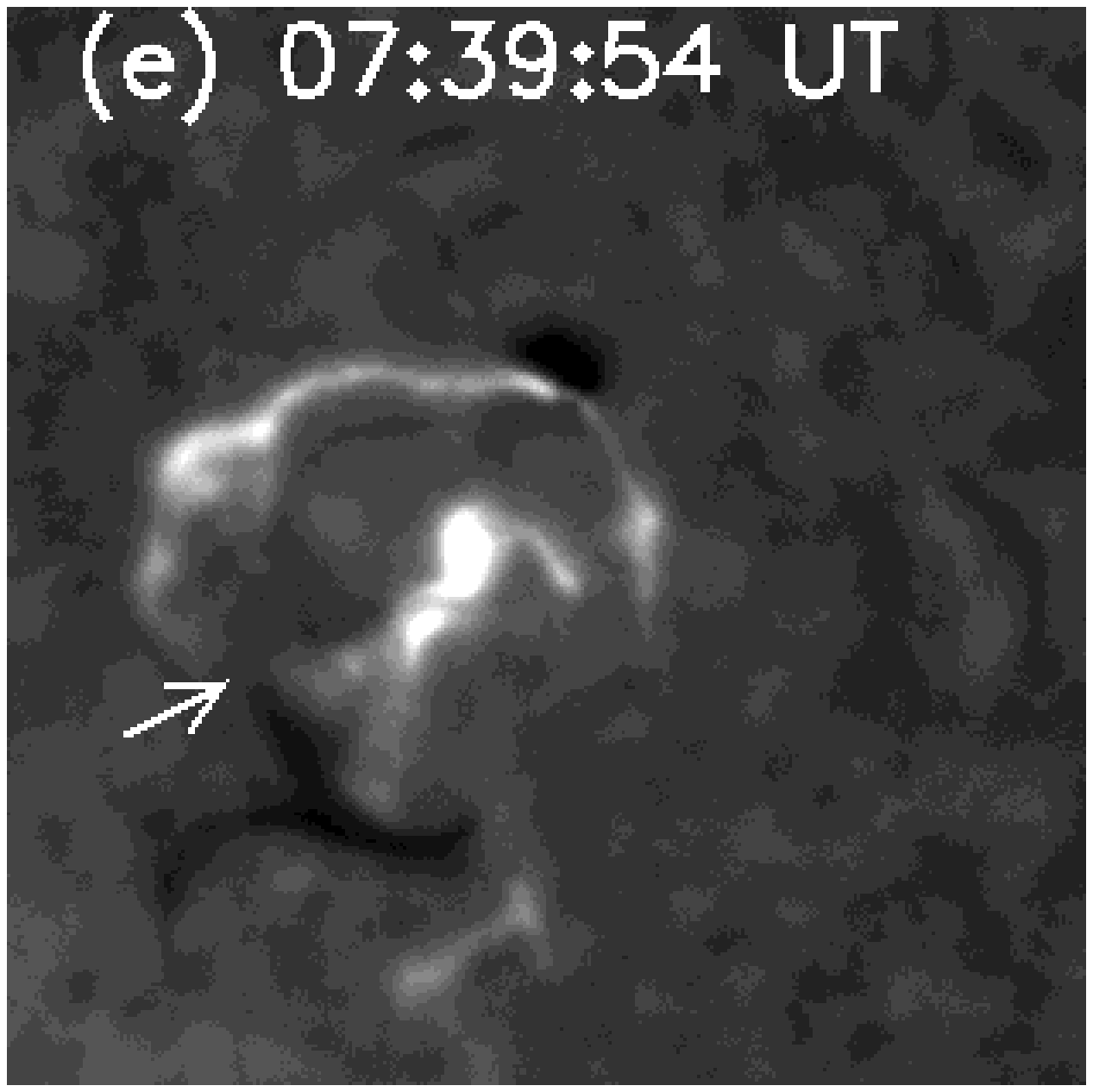}
\includegraphics[width=2.00cm]{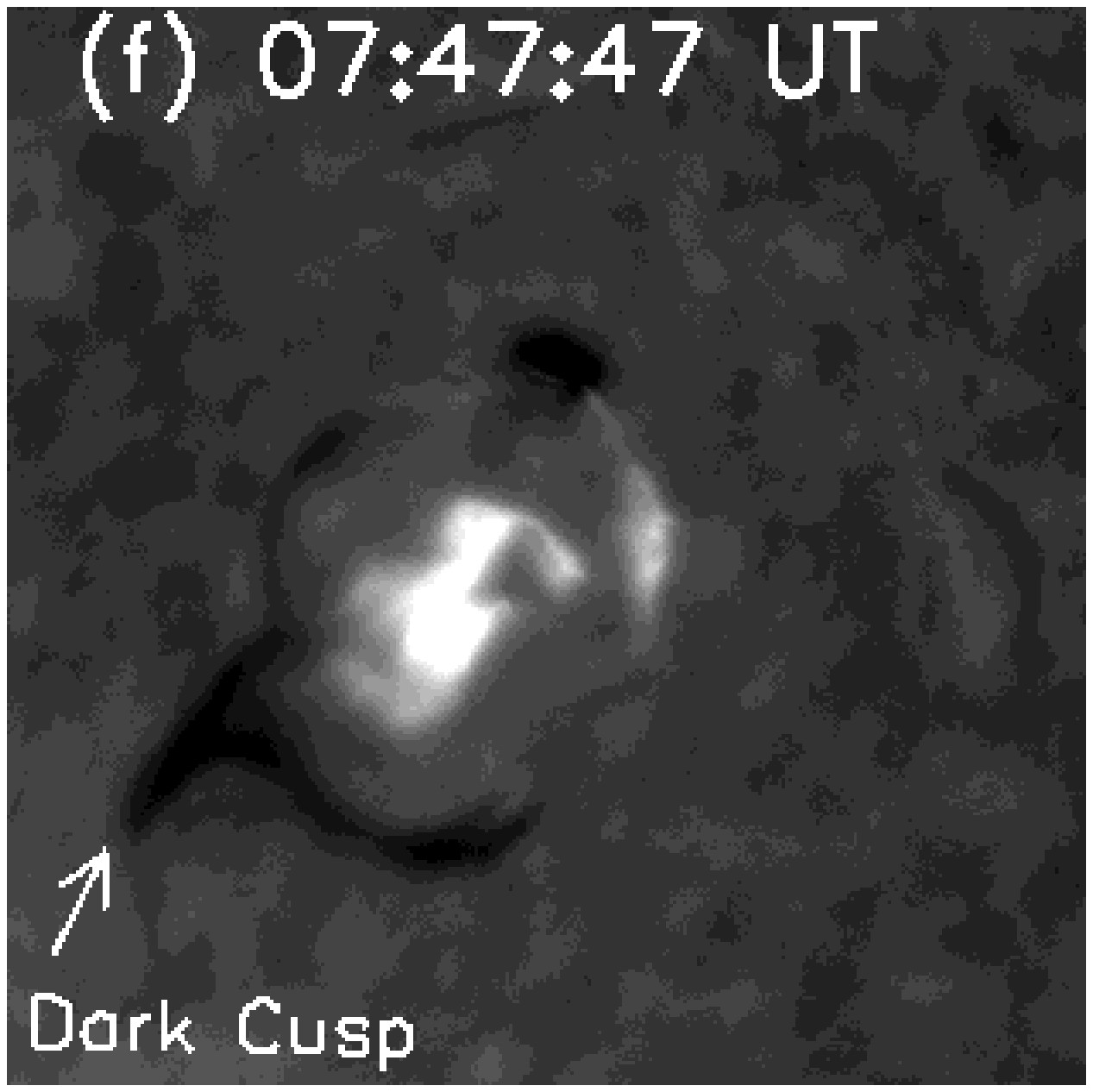}
\includegraphics[width=2.00cm]{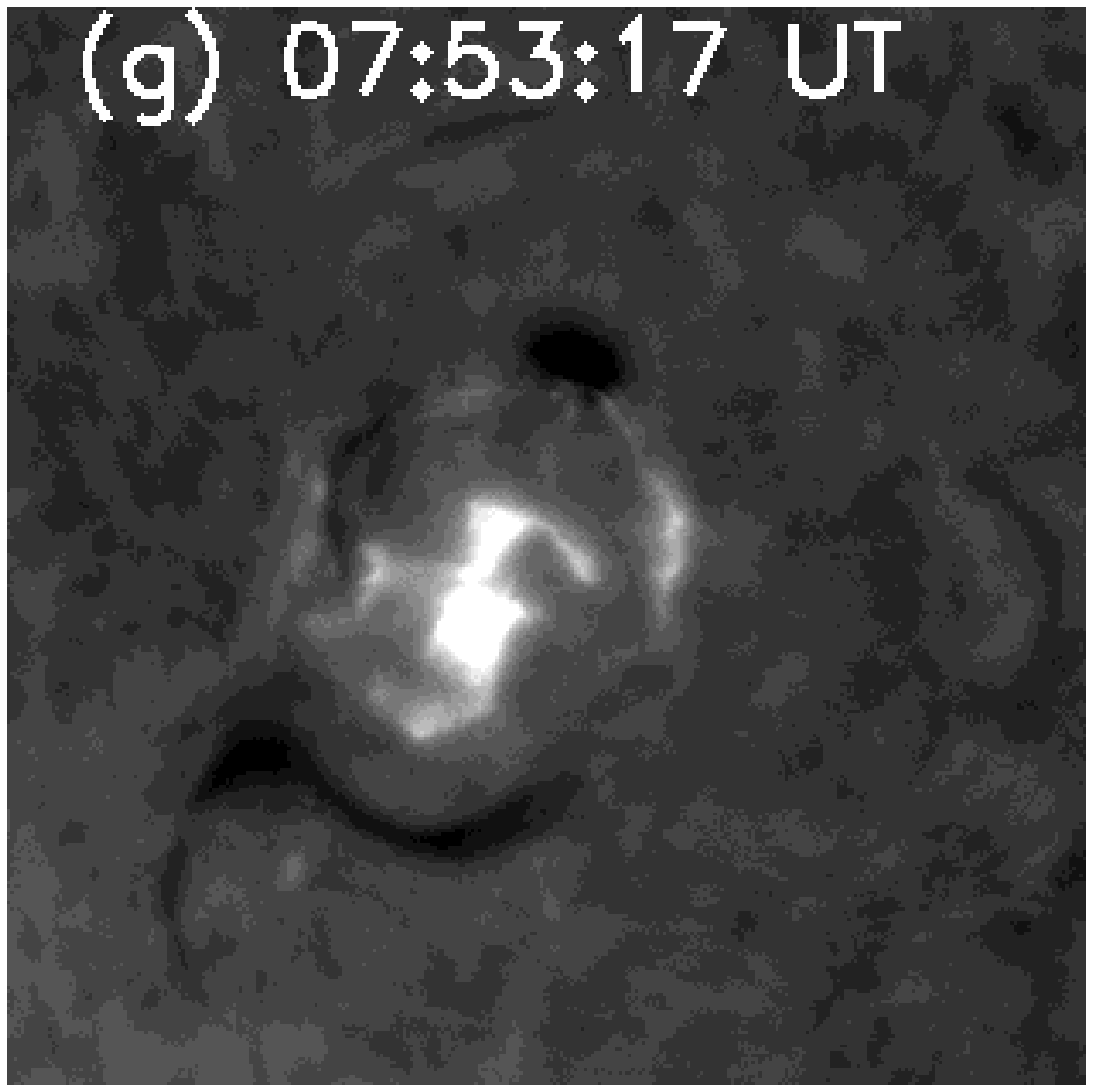}
\includegraphics[width=2.00cm]{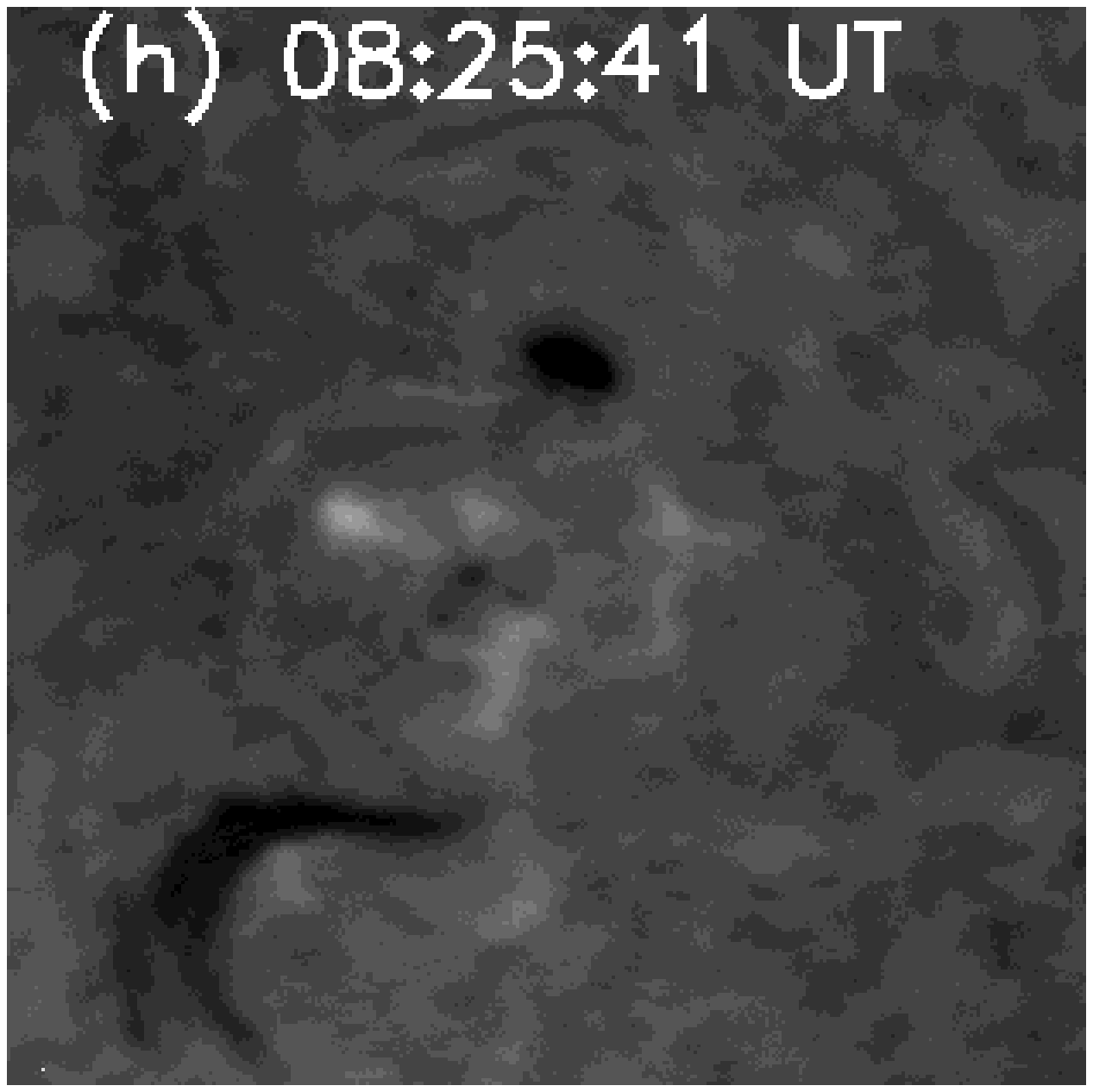}
\caption[]{\label{kumar-fig:2}
MDI magnetograms (top row) and H$\alpha$ images (middle and bottom
rows) during the flare events.}
\end{figure}


During 02:40--07:30~UT there was no significant activity at the flare
site (Fig.~1). However, the energy buildup from 01:30 to 08:00~UT is
revealed by the rotation of two sunspots of opposite polarity (the
northern spot moving clockwise, the southern one anticlockwise) at the
flare site (see arrows in the MDI images). During the onset of the
second event at about 07:35~UT, the twisted filament system, with one
footpoint attached to the sunspot group, goes through heavy
destabilization and the filaments approach each other similarly as in
the first event. The associated brightening, observed near the
filament, suggests that reconnection occurred between twisted field
lines and nearby small-scale fields. Furthermore, as the merging of the
filaments progresses (Fig.~2 d--h), the flare brightness rose to maximum 
intensity (Fig.~1). The H$\alpha$ images also revealed 
restructuring of the system after the flare maximum.

\begin{figure}
\centering
\includegraphics[width=5.3cm]{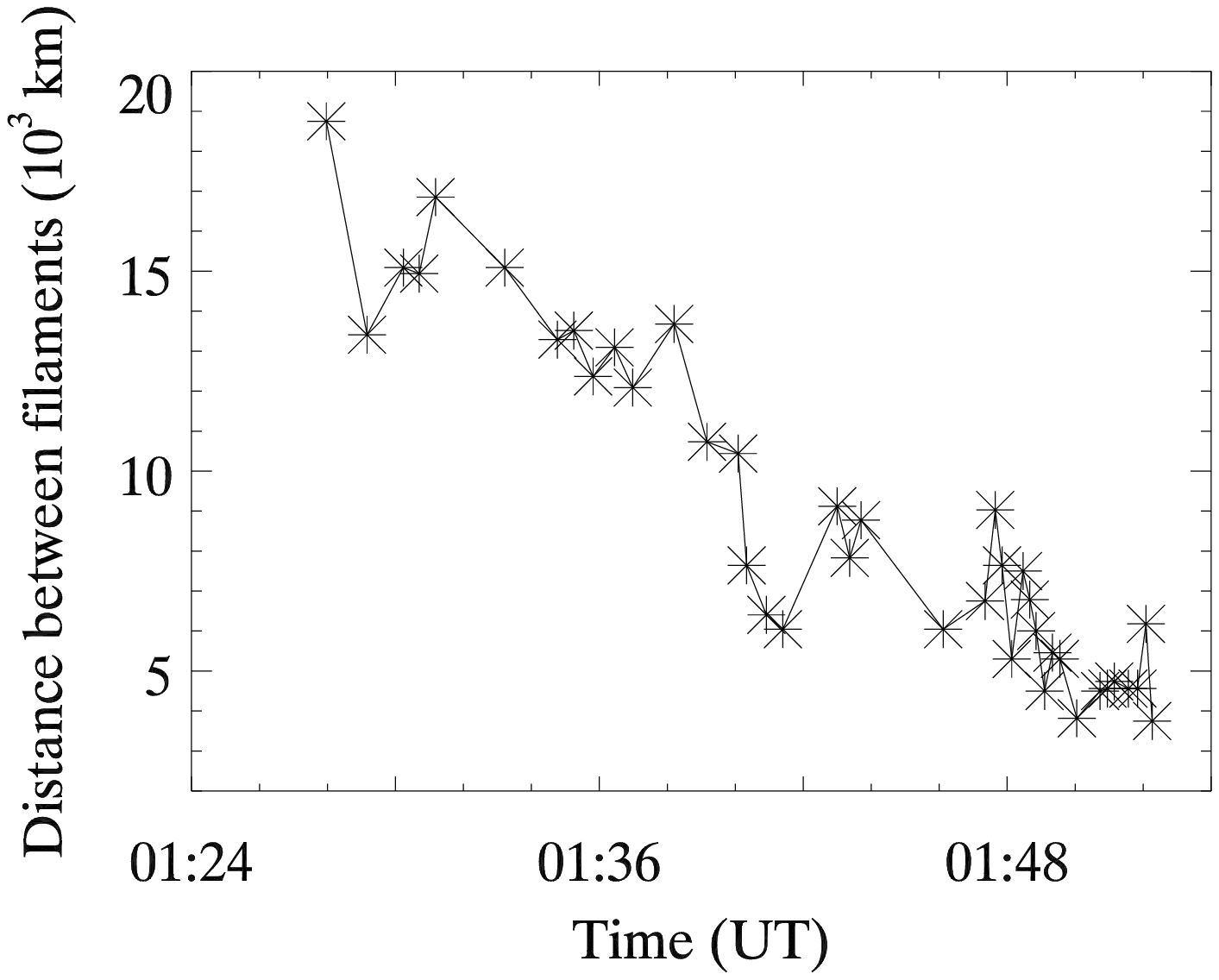}
\includegraphics[width=5.5cm]{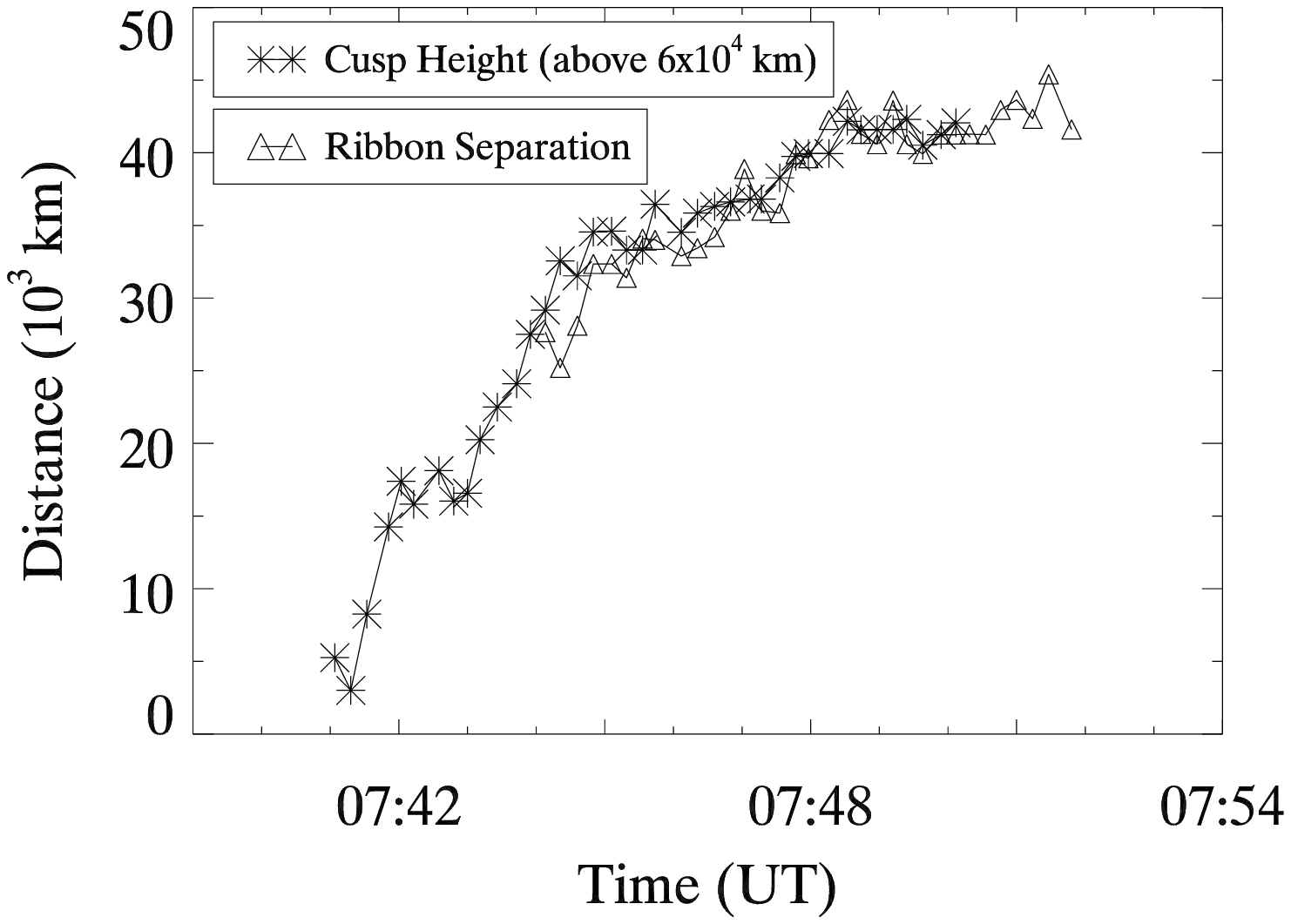}
\caption[]{\label{kumar-fig:3}
Left: separation of the two filaments, showing that they 
approached each other.  Right: cusp height variation and ribbon
separation during the flare event.}
\end{figure}

A dark cusp-shaped structure formed after the flare maximum (Fig.~2 f)
and started to move upward.  Its speed component in the plane of the
disk was about 5--6~\kms\ during 07:45--07:50~UT (Fig.~3). At
07:51~UT, the cusp erupted and some part of it fell back.  The
interesting point is that the brightening observed along the flare
ribbons and the rate of separation between the ribbons correlate well
with the eruption rate of the cusp (Fig.~3). The high correlation
(about 93\%) suggests that the magnetic reconnection and the rise of
the filament system played a prime role in the initiation of the
coronal mass ejection (CME). After the flare, the filament system
returned to its original, relaxed state.

Fig.~4 shows EIT and LASCO images taken at the time of the CME onset.
The CME was tracked further out using the interplanetary scintillation
(IPS) technique, which shows the geometry of the CME while crossing
the Earth's magnetosphere.


\begin{figure}
\centering
\includegraphics[width=2.6cm]{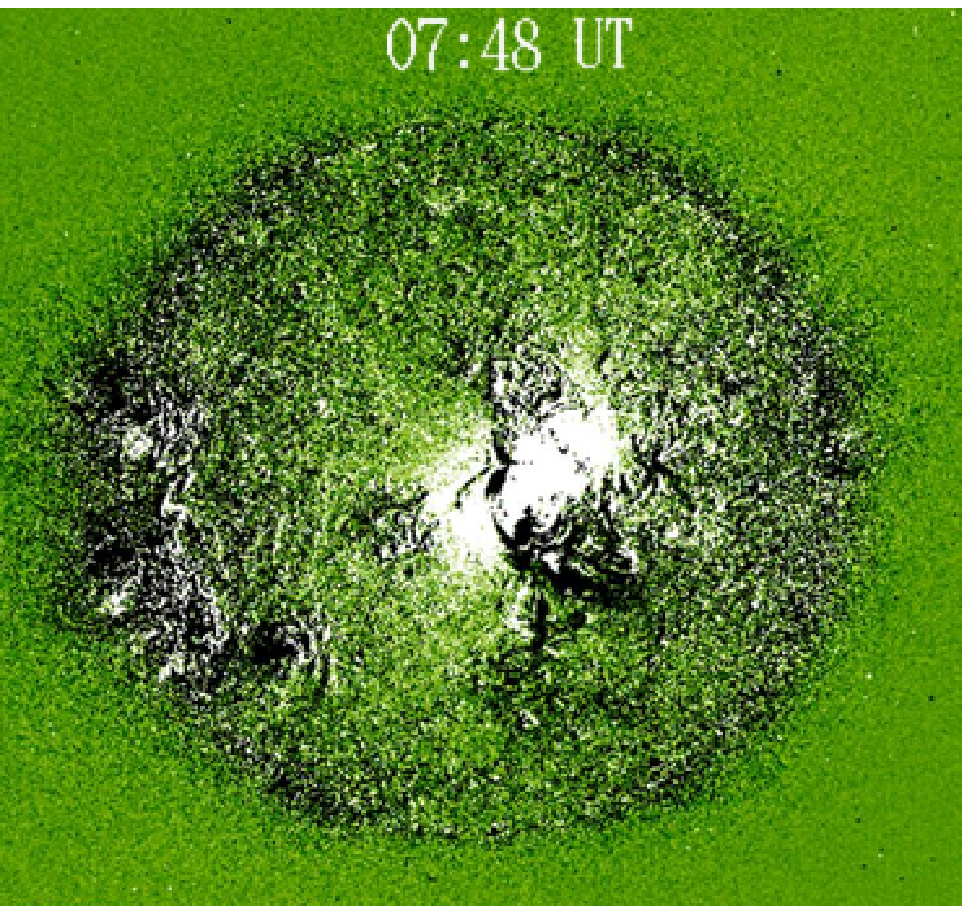}
\includegraphics[width=2.6cm]{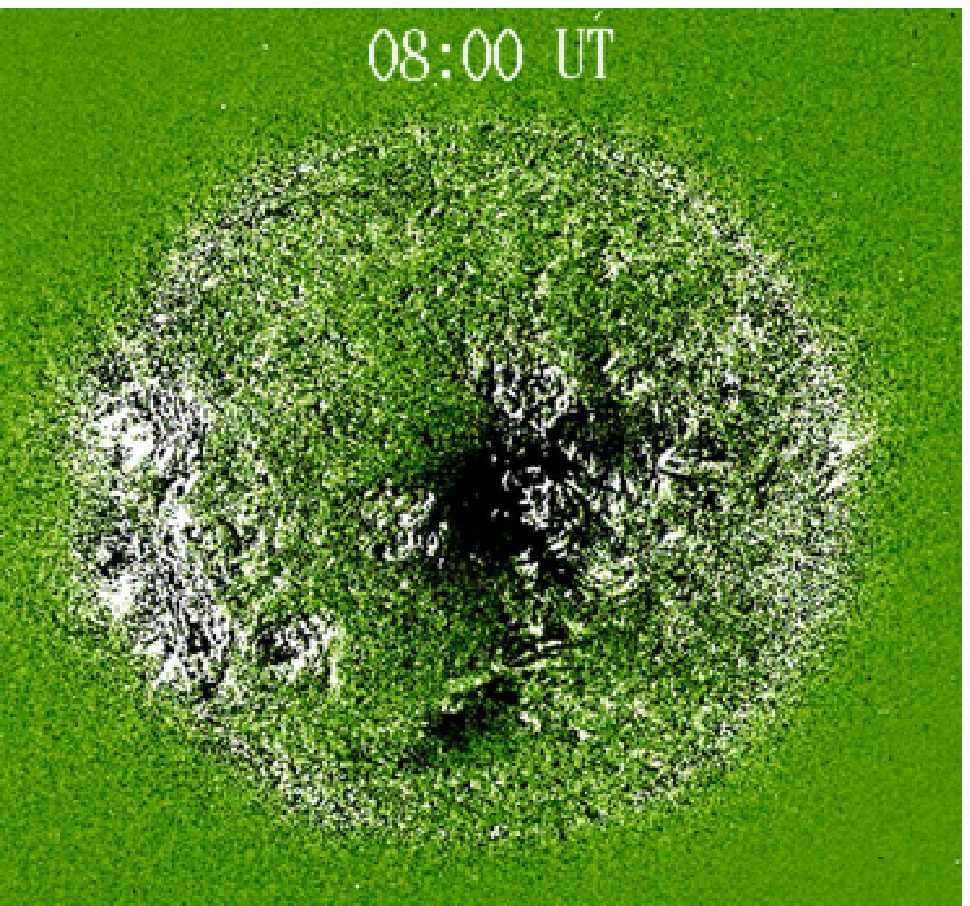}
\includegraphics[width=2.55cm]{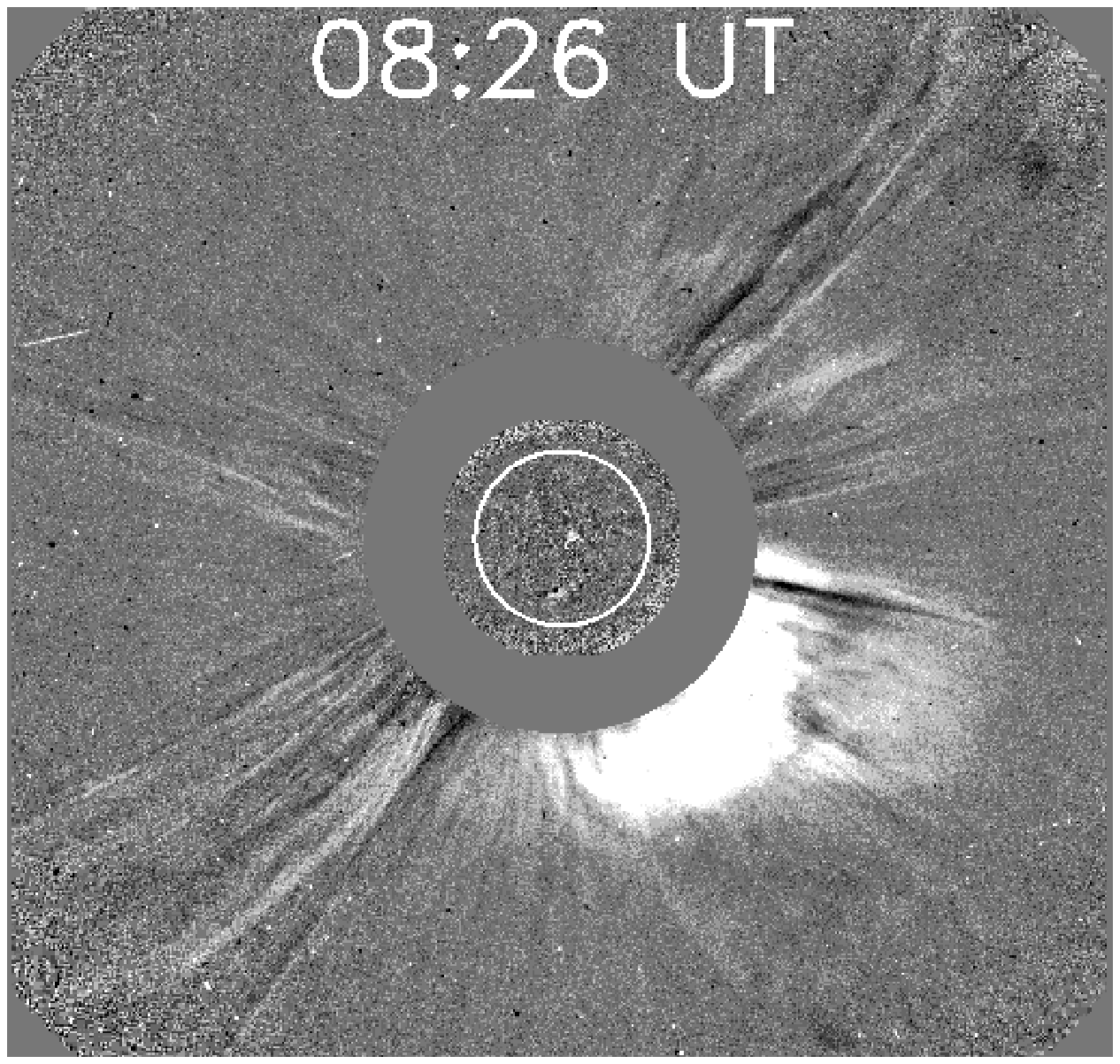}
\includegraphics[width=2.55cm]{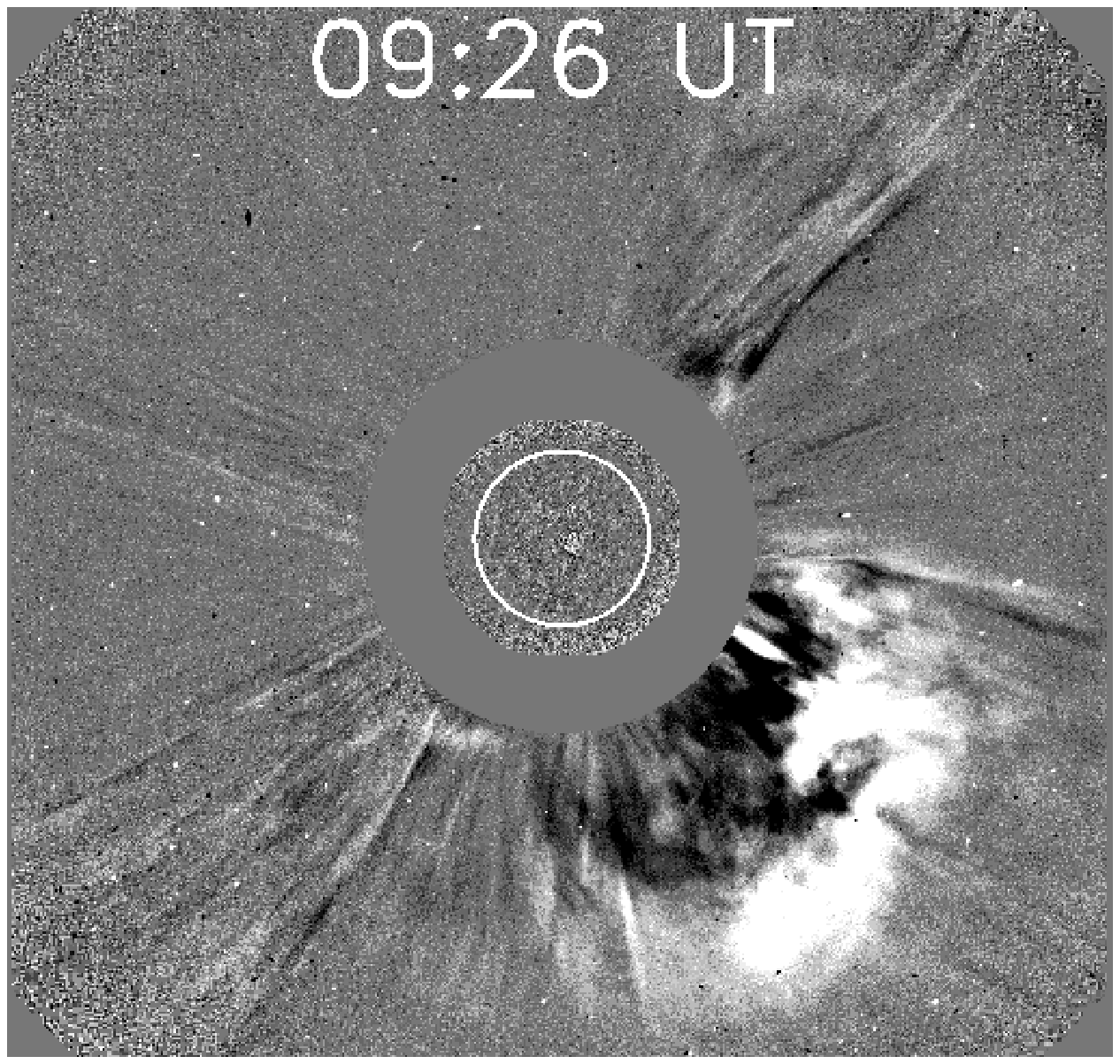}
\includegraphics[width=11.0cm]{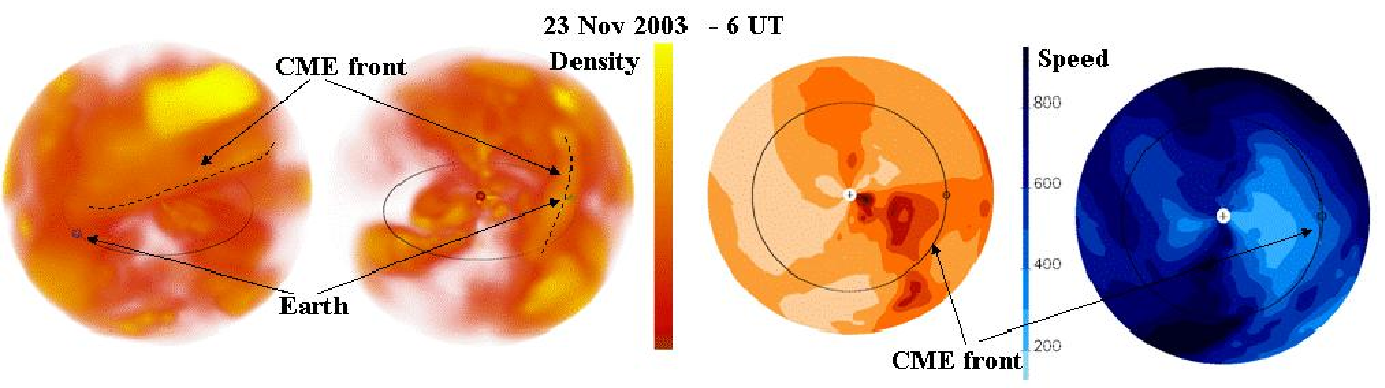}
\caption []{\label{kumar-fig:4}
Top: EIT difference images of the flare, showing the flare maximum at
7:48~UT and coronal dimming (left two images) and difference images of
the CME from the C2 coronagraphs on SOHO/LASCO, showing the CME association
with the second, M9.6 flare (right two images). Bottom: 3D view of
the CME in Ooty IPS images, showing the orientation of the flux rope
of about 70\deg\ with respect to the ecliptic plane.}
\end{figure}

\section{Results and discussion}                \label{kumar-sec: Results and Discussion}
This multi-wavelength study provides evidence that opposite
rotation of opposite polarity regions plays a crucial role in building
the magnetic energy required for the flare process. Sunspot rotation
is the primary driver of helicity production and injection into the
corona (\cite{kumar-2006SoPh..233...29T},
\cite{kumar-2002ESASP.477...47V}). Newly emerging flux plays a major
role in the destabilization of filaments. The cusp shape suggests the
formation of a magnetic null point in the high corona
(\cite{kumar-2003ApJ...592..597M}). The correlation between the
separation of flare ribbons and the expansion of the cusp structure indicates
that large-scale reconnection and particle acceleration occurred during
the cusp eruption.  The IPS technique shows that the flux rope is
oriented about 70\deg\ with respect to the ecliptic
plane. Therefore, in spite of a strong shock, the Earth-directed CME
caused only a moderate storm (Dst -85\,nT) at the Earth.

\begin{acknowledgement}
SOHO (EIT, LASCO, and MDI images) is a project of international cooperation
between ESA and NASA. PKM acknowledges the partial support for this study
by CAWSES-India Program, which is sponsored by ISRO.
\end{acknowledgement}

\begin{small}

\end{small}
\end{document}